# Different Algorithms *(Might)* Uncover Different Patterns: A Brain-Age Prediction Case Study

Tobias Ettling[†]
*Department of Computer Science*
*University of St.Gallen*
St. Gallen, Switzerland
tobias.ettling@gmx.de

Sari Saba-Sadiya[†]
*Informatik und Mathematik*
*Goethe-Universität*
Frankfurt, Germany
ORCID: 0009-0005-7482-3274

Gemma Roig
*Informatik und Mathematik*
*Goethe-Universität*
Frankfurt, Germany
roig@cs.uni-frankfurt.de

*Abstract*—Machine learning is a rapidly evolving field with a wide range of applications, including biological signal analysis, where novel algorithms often improve the state-of-the-art. However, robustness to algorithmic variability - measured by different algorithms, consistently uncovering similar findings - is seldom explored. In this paper we investigate whether established hypotheses in brain-age prediction from EEG research validate across algorithms. First, we surveyed literature and identified various features known to be informative for brain-age prediction. We employed diverse feature extraction techniques, processing steps, and models, and utilized the interpretative power of *SHapley Additive exPlanations* (SHAP) values to align our findings with the existing research in the field. Few of our models achieved state-of-the-art performance on the specific data-set we utilized. Moreover, analysis demonstrated that while most models do uncover similar patterns in the EEG signals, some variability could still be observed. Finally, a few prominent findings could only be validated using specific models. We conclude by suggesting remedies to the potential implications of this lack of robustness to model variability.

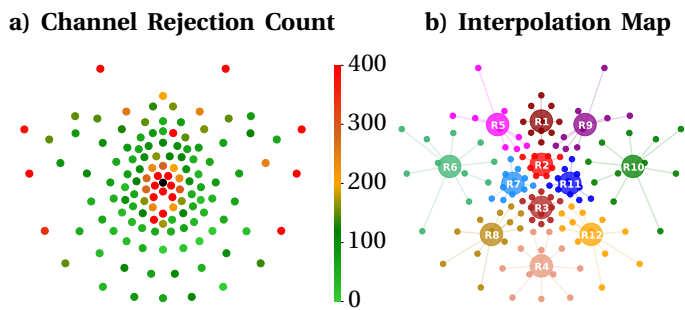

Fig. 1: Electrode positions as viewed from top (facing up). a) colors indicate (subject average) channel rejection counts across 3200 recordings. Dark-red electrodes (rejected over 400 times) were dropped. b) colors indicate interpolation groups. Electrodes in each of the 12 groups were interpolated into a single channel (denoted by large circles).

## I. INTRODUCTION

Machine learning is ubiquitous in biological signal research and can be useful for identifying the features carrying information relevant to a specific task. Yet, there is a lack of guidelines towards algorithm selection, and researchers seldom cross-validate their hypotheses across diverse algorithms. This is especially an issue as researchers often pick only the best performing algorithm for further analysis (for instance [1]–[4]). In this paper we formalized a framework for investigating agreement between the feature importance ranking of different algorithms, and explored whether various previous findings regarding the relevance of specific features for brain-age prediction can be consistently replicated. Analysis uncovered mild inconsistencies across algorithms, demonstrating that different algorithm types (regression based, decision tree based ...) may uncover different patterns for the same data. Code and data necessary to reproduce our results were made available [1].

The contributions towards this paper are as follows:
- We design a feature extraction and model training pipeline that achieves state-of-the-art performance in Brain-Age prediction (section V).
- We formalize an extension to SHAP values that can be used to explore model agreement (VII).
- We identify 16 established findings in brain-age prediction literature (section II), and use our model interpretation framework (the *SHAPagreement*) to examine whether they can be replicated by each of 10 commonly used algorithms (section VIII).

## II. RELATED WORK

The changes that occur in our brains as we age are subject to intensive research. In this section we provide an overview of previous studies and summarize their results in Table I. These studies used various algorithms. We will attempt to replicate each of these findings in subsequent sections (sections VII and VIII).

Babiloni et al. [6] investigated linear and non-linear correlations between the EEG signal features and chronological age. By contrasting the EEG recordings from two distinct subject groups — one aged 18 to 50 and the other 51 to 85 years — they revealed that delta and alpha magnitudes in parietal, occipital, temporal, and limbic areas were more pronounced in younger individuals in comparison to elderly. Moreover, they discovered that activity sources, while exhibiting linear correlations, also displayed intriguing non-linear relationships with age. Despite the statistically significant results (as low as $p = 0.00003$) reported, the correlation between age and EEG features was relatively modest, ranging from $r = 0.24$ to $r = 0.43$. These results indicate that solely analyzing individual features of the brain's activity pattern, as opposed to integrating them using machine learning models, may not provide a comprehensive understanding of how aging impacts the brain [6]. Inspired by this, we focus on machine learning for brain-age prediction research.

---
[†]Authors contributed equally to this work.
[1]https://github.com/Arsu-Lab/Different-Algorithms-Uncover-Different-Patterns-BrainAge-Prediction

TABLE I: Prior research findings related to changes in EEG patterns during maturation. Age in Years: the age range used. Study Type: specifies if the study was cross sectional or longitudinal.

| Frequency Band | Findings | Age in Years | Study Type | Paper |
|---|---|---|---|---|
| delta | Central delta features show significant importance to random forest for predicting the age. | 5 to 18 | longitudinal | [1] |
| | Absolute delta activity decreased with age. | 8 to 12 | cross-sectional | [5] |
| | Occipital delta magnitude was higher for younger subjects (18 to 50 years) than for older subjects (51 to 85 years). | 18 to 85 | cross-sectional | [6] |
| | Occipital delta magnitude correlates linearly with the subject age. | | | |
| theta | Temporal theta features show significant importance to random forest for predicting the age. | 5 to 18 | longitudinal | [1] |
| | Theta power decreases with age. | 9 to 16 | cross-sectional | [7] |
| | | 4 to 17 | longitudinal | [8] |
| alpha | Increase in alpha activity beginning in posterior regions and ending in anterior regions. | 5 to 18 | longitudinal | [1] |
| | Peak alpha frequency increases with age. | | | |
| | Frontal lower and parietal alpha features show importance for predicting the age with random forest. | | | |
| | Alpha waves show a non linear pattern with ageing. | 18 to 85 | cross-sectional | [6] |
| | Relative alpha activity increases with age. | 8 to 12 | cross-sectional | [5] |
| beta | Beta frequency power shows importance for predicting the age with random forest. | 5 to 18 | longitudinal | [1] |
| | Spectral flatness of beta band was most important for model predictions. | 18 to 58 | cross-sectional | [2] |
| | Beta band power positively correlated with age . | 8 to 12 | cross-sectional | [5] |
| whole spectrum | PSD slope was more negative in pre-teen (under age 13) vs teen subjects (age 13 to 16). | 9 to 16 | cross-sectional | [7] |
| | Multiscale entropy increased in frontal and central regions with ageing. | | | |

Machine learning models are remarkably capable of capturing complex patterns from massive data-sets. In recent years, the push for model clarity has birthed cutting-edge techniques like feature importance methods. These analytical methods center on models that utilize handcrafted features from EEG signals, spanning statistical, time-domain, and spectral dimensions (see Table III). Feature extraction refines the analysis and reduces data complexity. Subsequently, feature importance techniques assess the relevance of these extracted features for predictive tasks.

Al Zoubi et al. [2] assessed various machine learning models for brain age prediction using handcrafted EEG features from subjects aged 18 to 58. Beyond random forest and support vector machines, they introduced a stacking (ensemble learning) approach. This method yielded the top performance with an MAE of 6.87 years ($R^2 = 0.37$). Feature evaluation highlighted the spectral flatness (Wiener entropy) in the beta band, particularly from temporal parietal channels, as a dominant influence in age prediction, followed by relative beta band power and alpha entropy. In another study by van Noordt and Willoughby [7], the spectral slope feature derived from the Power Spectral Density (PSD) stood out as an informative indicator of brain maturation. A more negative PSD slope was observed in pre-teens (under 13) than in teens (13 to 16), indicating a shift from low to high frequency content as we mature. Collectively, these insights highlight the potential of features derived from EEG signal PSD in age prediction.

Another established finding was a decrease in power within lower frequency bands as we age [1], [5].

## III. DATA-SET

We used the NeuroTechX brain age prediction dataset [9]. The dataset consists of EEG recordings from 1200 subjects, including measurements of 40 seconds for eyes closed (EC) and 20 seconds for eyes open (EO) resting state. The EEG recordings were acquired from 129 channels, inclusive of a *Cz* reference channel, and were sampled at a frequency of 500 Hz. The age of the subjects ranged continuously from 5 to 22 years, with an average age of $\mu_{age} = 10.4$ years and a standard deviation of $\sigma_{age} = 3.5$ years. Processing was done with the MNE-Python [10] package. We adopted the *"GSN-HydroCel-129"* montage. We then resampled to 250 Hz, applied a zero-phase filter (high-pass at 50Hz, low-pass at 0.5 Hz), and removed the reference channel *Cz*.

### A. EEG Processing

We used the *autoreject.RANSAC* method from the *autoreject* package [11], [12], for channel rejection. First we split the data into 1 sec wide epochs with each containing 250 data points for every channel. Channels flagged in over 40% of these epochs are dropped. To make this determination within each epoch, we employ the RANSAC technique [13]. We randomly sampled 25% of the channels and interpolate them to estimate a representation of the other 75% of channels. The channel is flagged as an outlier if the correlation with the interpolated signal is lower than a given threshold (75%). These steps where repeated 50 times to find the best consensus set of channels for each epoch. We analysed the rejection frequency for each channel across all 3200 recordings (shown in Fig. 1 a)). Results unveiled a recurring rejection of frontal electrodes. These electrodes tend to capture muscle artifacts generated by the contraction of various face-muscles and are often noisy [14]. Additionally, electrodes situated in close proximity to the zero reference channel *Cz* also exhibited a notable rejection rate, likely due to strong attenuation when referencing due to the high correlation with the reference channel *Cz*.

TABLE II: The different training sets

| Training Set | Channels | State | Total Features |
|---|---|---|---|
| 128-All | 128 | EC, EO | 32256 |
| 128-EC | 128 | EC | 16128 |
| 128-EO | 128 | EO | 16128 |
| 12-All | 12 | EC, EO | 3024 |
| 12-EC | 12 | EC | 1512 |
| 12-EO | 12 | EO | 1512 |

TABLE III: Overview of measures selected to be extracted from the EEG signals

| | Measure Name | Description |
|---|---|---|
| **Temporal Features** | Mean | Mean of the signal in volt. |
| | Standard Deviation | The average difference between the signal and its mean in volt. |
| | Peak-To-Peak Amplitude | Distance between the maximum and minimum measure of the signal. |
| | Line Length | Physical length of the signal given as the sum of the absolute difference between consecutive time measures of the signal. |
| | Zero Crossings | How often the signal changes from positive to negative. |
| | Skewness | Quantifies the asymmetry of the signal. |
| | Kurtosis | Describes the "tailedness" of a distribution compared to the normal distribution. |
| | Hjorth Complexity | Measures the complexity or irregularity of the signal |
| **Frequency Features** (log-log PSD) | Intercept | Intercept of the regression line, fitted to the power spectral density (PSD) of the signal with the y-axis. |
| | Slope | The slope of the regression line. |
| | Mean Square Error | The mean squared error between the regression line and the log-log power spectral density (PSD) of the signal. |
| | R2 coefficient | Quantifies how well the regression line fits the log-log power spectral density (PSD) of the signal. |
| | Band Power ($\delta, \theta, \alpha, \beta, \omega$) | The power of the signal in a frequency band. |
| | Wavelet Coefficient Energy ($\delta, \theta, \alpha, \beta, \omega$) | The energy of wavelet decomposition coefficients. |
| | Hjorth Complexity | Measures the complexity or irregularity of the signal's power spectrum. |
| **Statistical Features** | Quantile (5%, 25%, 75%, 95%) | Describes the distribution of the signal for percentiles. |
| | Higuchi Fractal Dimension | Quantifies the fractal complexity or self-similarity of the signal. |
| | Sample & Approximate Entropy | Quantifies the complexity or regularity of the signal. |
| | Spectral Entropy | Is the Shannon entropy of the signal's power spectrum. |
| | SVD Fisher Information | Singular Value Decomposition (SVD) Fisher Information per channel |
| | Hurst Exponent | Characterizes the long-term memory or self-similarity of the signal. |

These results are within expectation, and the prepossessing should not influence our main analysis. On average $\mu_{rej} = 9$ channels were rejected per recording (standard deviation of $\sigma_{rej} = 13$). As a filtering step, recordings with over 30 rejected channels (115 recordings) were excluded from further analysis.

### B. Channel Interpolation

We segmented all channels into 12 distinct regions, each consisting of 9 to 12 channels, as depicted in Figure 1. For the interpolation we use the *combine_channels()* method from the MNE package [15], resulting in a single representative signal for each individual region. We will later compare the prediction performance, with and without regional channel interpolation.

### C. Eyes Open / Closed Data-sets

Different tasks recruit different cognitive mechanisms. Therefore, for a full picture of how aging impacts cognitive processing we prepared 3 data-sets. This was achieved by extracting the features from only EO (eyes open), only EC (eyes closed) or both resting states. Furthermore we ran the analysis with and without channel dimension reduction. This resulted in 6 total data-sets. In '128-' sets, features were extracted from all 128 channels, excluding the 'Cz' channel, encompassing both EO and EC features. In '12-' sets, we applied regional channel interpolation to condense the channel dimension to 12 regional channels, from which we extracted both EO and EC features. These two primary sets were then varied to only include EO (128-EC, 12-EC) and EC features (128-EC, 12-EC). For all sets we used the same methods for feature extraction only varying the recording state (EO, EC) and channel dimension. All training samples were flattened to one-dimensional vectors that are uniquely associated with the channel, frequency band, recording state, and extracted features from which they originated. In Table II you can see an overview of these sets with corresponding feature dimensions. Notably the dimensions are by far smaller when applying the regional interpolation strategy, see 128-All, 128-EO, 128-EC.

## IV. FEATURE SELECTION

We selected the features for our study based on those used in the prior brain-age research we surveyed, mainly from [2], [7], [17]. All of our features were extracted using the *extract_features* method from the *mne_features* python library [15]. Our set of features included the following time domain features: the mean, standard deviation, peak-to-peak amplitude, line length [18], zero crossings, skewness, kurtosis and hjorth complexity parameter [19]. For frequency domain features we used the *compute_spect_slope* method [20], [21] to compute the intercept, slope, mean squared error and R2 coefficient of a regression line fitted to the logarithmic transformed PSD of the signal [22]. Furthermore we used the frequency band power and hjorth complexity parameter specific to the frequency domain [23]–[25]. This includes the following statistical features: the quantile, higuchi fractal dimension [19], [25], [26], sample and approximate entropy [27] of the signal and the entropy of the power spectrum [28]. To measure signal consistency we added the hurst exponent [29] as a higher order feature. In Table III you can find an overview and description of all extracted features.

### A. Feature Extraction

The feature extraction pipeline used in our work closely resembles the framework outlined in previous brain-age prediction research [2], [17].

Features were extracted in each frequency band ($\delta = 0.5 - 4Hz$, $\theta = 4 - 7Hz$, $\alpha = 7 - 14Hz$, $\beta = 14 - 30Hz$), as well as the whole spectrum ($\omega = 0.5 - 30Hz$). First, recordings were band-pass filtered yieldings 5 recordings per subject and divided into 10 epochs (2 second epochs for eyes open, and 4 seconds for eyes closed). After extracting the features for each epoch and frequency band we averaged the feature values across the epochs resulting in a vector of size [5 Frequency bands × Number of Channels × Number of Features] for each recording of the original dataset. See supplementary section I for more details.

TABLE IV: An overview of the models we evaluated. These models were trained on the extracted feature-sets with the primary aim of predicting the chronological age of each subject.

| | Model Name | Description |
|---|---|---|
| Regression | Elastic Net Regression | Is a regression technique and uses a combination of L1 and L2 penalty, effectively performing feature selection on correlated features by shrinking their coefficients. |
| Regression | Lasso Regression | Is a regression technique and only applies the L1 penalty to the loss function. |
| Regression | Kernel Ridge Regression | In Ridge Regression, the penalty term is the L2 norm (squared magnitude) of the coefficient vector. This model makes use of the so called "kernel trick", which is used to capture the non-linear similarities between data points. |
| Tree-Based / Boosting | XGBoost | This model is build on the gradient boosting architecture and therefore is a ensemble method. At the core of this framework are multiple decision trees (weak learners), each trying to correct the errors made by the previous tree. XGBoost implements L1 and L2 regularization. |
| Tree-Based / Boosting | CatBoost | This model is build on the gradient boosting architecture and therefore is an ensemble method. The main difference to XGBoost is the automated category encoding of the features (categorical, missing, unprocessed features). Another core difference is the method used to prevent overfitting, named "ordered boosting" [16]. |
| Tree-Based / Bagging | Random Forest | As the name implies, the model consists of training multiple decision trees using the bagging (bootstrap and aggregating) strategy. It therefore belongs to the ensemble learning methods. The output of each weak learner is combined during prediction, which improves performance and reduces overfitting. |
| Other / Bagging | Bagged K-Nearest Neighbors | Implements the bagging (bootstrap and aggregating) strategy in extend to the KNN architecture and therefore belongs to the ensemble learning methods. The bagged version of KNN aims to enhance the performance of the standard KNN model. |
| Other | K-Nearest Neighbors | This model underlies a completely different architecture. Predictions are made upon data points that are in close distance (neighbor) to each other. The majority vote of all classes assigned to data points in proximity are used to predict the outcome of a new data point. This method works for classification and regression. |
| Other | Support Vector Machine | SVM calculate hyperplane, separating data points with different classes while maximizing a loss function. To handle non-linear boundaries (complex patterns), it uses the so called "kernel trick". The "trick" replaces the dot product with a kernel function, which captures non-linear similarities between data points. |
| Other | Multi-Layer Perceptron | Is a type of feed-forward neural network, implementing multiple hidden layers between the input and output layer, making it a deep learning model. It can be used for regression tasks but also for classification by assigning probabilities to each class (node) of the output layer. |

## V. MODELS

We trained ten different machine learning models on the age-prediction task (see Table IV). All models were trained with a stratified 3-fold cross validation, implemented with the *scikit-learn* package [30]. While the original continuous ages were used during model training, the stratified folds were created using discretized ages by year. Additionally, we applied standardization across the time axis to the data for all non-tree-based models. For all tree based models, RandomForest, XGBoost, and CatBoost we label encoded the target values. The evaluation metric we used was the mean absolute error (MAE) for each fold. The overall score for each model is the mean across the fold scores. We used the bayesian hyper-parameter tuning method *BayesSearchCV* from *scikit-optimize* [31].

## VI. BRAIN AGE PREDICTION RESULTS

Despite the relief in computational cost for 12-All in comparison to 128-All, the difference in prediction performance was modest. For XGBoost, CatBoost, RandomForest, ElasticNet, SVR, and Lasso Regression the performance was slightly higher when training on 128-All. In contrast, the opposite was true for the K-Nearest Neighbours, Bagged KNN, KernelRidge and Multi-Layer Perceptron models. For a comprehensive list of model scores refer to Table V.

TABLE V: The average fold score (mean ± std) across all folds for all models on the different dataset variations. XGBoost always had the best performance. The best score for each model is bolded.

| Model | 12-All | 12-EC | 12-EO | 128-All | 128-EC | 128-EO |
|---|---|---|---|---|---|---|
| MLP | 2.04±.08 | **2.01**±.03 | 2.43±.01 | 2.35±.06 | 2.43±.05 | 2.28±.07 |
| KernelRidge | 1.95±.03 | **1.92**±.05 | 2.12±.01 | 2.15±.03 | 2.11±.06 | 2.27±.1 |
| KNN | 1.96±.08 | **1.93**±.07 | 2.09±.09 | 1.99±.08 | 1.98±.06 | 2.12±.04 |
| BaggedKNN | 1.93±.06 | **1.9**±.07 | 2.09±.06 | 1.97±.08 | 1.94±.06 | 2.11±.07 |
| Lasso | 1.92±.07 | 1.94±.07 | 2.13±.03 | **1.86**±.04 | 1.93±.06 | 2.05±.05 |
| EleasticNet | 1.92±.06 | 1.94±.06 | 2.11±.02 | **1.84**±.04 | 1.91±.03 | 2.01±.07 |
| SVRegression | 1.74±.06 | 1.77±.12 | 1.94±.04 | **1.73**±.05 | 1.76±.07 | 1.9±.04 |
| RandomForrest | 1.77±.05 | 1.77±.05 | 1.9±.06 | 1.73±.06 | **1.72**±.05 | 1.88±.07 |
| CatBoost | 1.69±.02 | 1.73±.04 | 1.81±.04 | 1.67±.03 | **1.67**±.01 | 1.81±.05 |
| XGBoost | 1.68±.01 | 1.7±.04 | 1.8±.01 | **1.62**±.03 | 1.65±.03 | 1.78±.05 |

Best results were obtained using both eyes open (EO) and eyes closed (EC) features, with EC-only close behind. EO-only training lagged in performance. This contrasts Khayretdinova et al.'s findings [32], where their CNN model's performance was fairly consistent across EC or EO data. This suggests our feature extraction may overlook fine grained information in the EO-EEG.

The variations in model performance for KNN, BKNN, KernalRidge and MLP correspond to the respective feature dimensions of the training data. Sets containing only EC or EO features exhibit half the feature dimensions compared to the set encompassing both state features. Additionally, in 12-All, the feature dimensions were reduced by a factor of 10.66 compared to 128-All. The consistent under-performance of KNN, BKNN, KernelRidge, and MLP on high-dimensional data underscores their susceptibility to the challenges posed by the curse of dimensionality [33]. Conversely, Lasso, ElasticNet, SVR, Random Forest, CatBoost, and XGBoost emerged as the top-performing models on the high-dimensional dataset encompassing all channels, reaffirming their robustness in handling high-dimensional data when trained correctly.

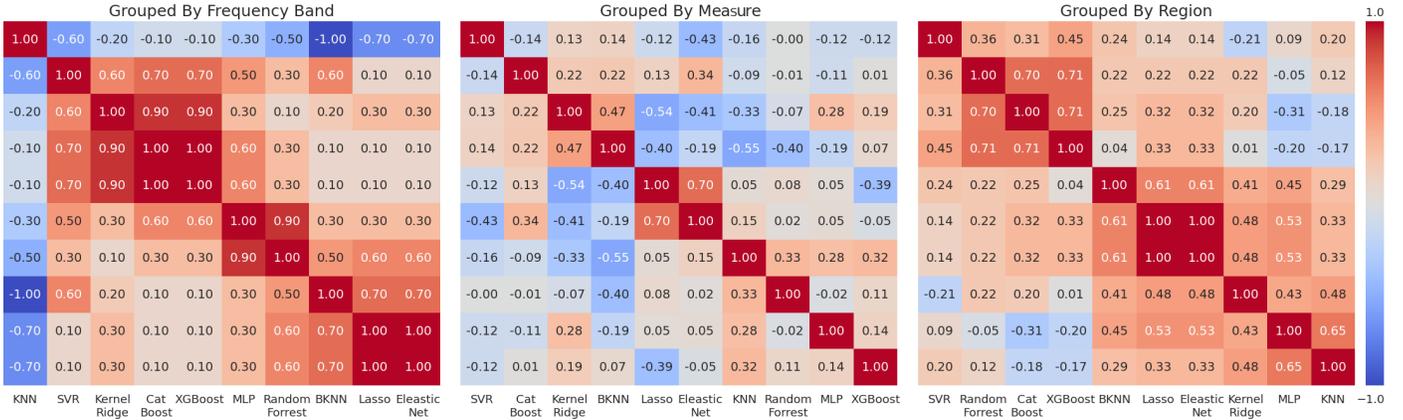

Fig. 2: Heatmaps illustrating the *SHAPAgreement* (rank order correlations of commutative shapley values) among features, grouped by (1) the frequency bands, (2) feature measure, and (3) electrode regions. The ranking reflects their importance to the respective model. SHAP values were computed for unseen samples with models trained on 12-All.

Our best performing model, XGBoost, achieved results comparable to the top-three models that participated in the Brain Age Prediction Challenge [9]. However, we learned that at least two of the top-3 models were trained on additional data that happened to included the hidden test set. Therefore, we believe that our models are capable of matching the performance of state-of-the-art algorithms when trained on comparable data.

## VII. MODEL INTERPRETATION

SHapley Additive exPlanations (SHAP) is a method used to quantify the importance of a feature in the model decisions. The concept was introduced by Lloyd Shapley's in game theory. This metric assigns scores to features based on their contribution to prediction outcome [34]. Despite SHAP values having a number of limitations that complicate their application to EEG data - for instance, EEG channels and extracted features are highly correlated - they have gained popularity in EEG research [35]. Our work utilized the SHAP python packadge [36] to calculate the SHAP values for our EEG features given the pre-trained age prediction model. Leveraging SHAP's additive properties, we summed up the absolute values of these features to determine their collective contribution.

$$\phi_{x'}(m) = \sum_{x_i \in x'} |\phi_{x_i}(m)|$$

Where $x_i$ is the input feature, $m$ is the trained model, and $\phi_{x_i}(m)$ is the SHAP value of $x_i$ given $m$. These collective contributions are finally used to calculate the spearmann rank-order coefficients for every pair of algorithms, resulting in a *SHAPagreement* measure that quantifies algorithm agreement on feature importance (Fig. 2).

### A. Feature Importance Correlations Across models

We employed a variety of models with diverse architectures and strategies. To examine whether these models exhibit differing patterns in feature contribution for predictions we analyzed three feature groupings. First, we grouped features based on in which of the five frequency bands they originated (Fig. 3). Second, we grouped features by the 19 different measures (Table III) employed for their extraction, as sourced from the *mne_features* python library [10]. Third, we grouped by the features using the region of the signal they were obtained from (Fig. 1 b).

In this context, we computed SHAP values for all models trained on 12-All, where we reduced the channel dimension to 12 regional interpolated channels and included features for both eyes opened and closed. This decision was based on the stability of performance across models and the significantly lower computational cost. Resulting in three rankings of features by importance for each model (Fig. 3).

After computing the feature group importance for the two distinct groupings, we calculated the rank-order correlation between the importance ranking of these groups for each model. This allows us to construct a correlation matrix to visualize the agreement between models on which features and frequencies were most informative of brain-age (Fig. 2). Notably, CatBoost and XGBoost displayed a perfect positive correlation, likely attributable to both models employing tree architectures with gradient boosting. However, it's worth highlighting that CatBoost employs a custom technique called "ordered boosting" to mitigate overfitting [16], distinguishing it from XGBoost.

Additionally, the ElasticNet and LassoRegression models also exhibited a perfect positive correlation. The primary distinction between these models lies in their regularization techniques: ElasticNet combines both the Lasso (l1) and Ridge (l2) penalties, whereas the LassoRegression model solely incorporates the Lasso (l1) penalty. The low correlation of KernalRidge with ElasticNet and LassoRegression likely stems from the optimal hyper-parameter setting we found for the l1_ratio in the ElasticNet model, set at 0.8. This implies that the Lasso penalty was more useful for age-prediction than the Ridge penalty, perhaps due to harsher penalty for inclusion of task-irrelevant features.

The KNN model exhibited the weakest correlation with all other models (Fig. 2). KNN is an instance-based algorithm that relies on measuring distances between data points to make predictions. Interestingly, employing the bagging strategy to the KNN model (BaggedKNN) showcased a perfect negative correlation to the basic KNN, and more similar behavior to the other models.

The highest correlation for the measure importance rank order was between the LassoRegression and ElasticNet model followed by the correlation between the Kernal-Ridge and BaggedKNN model. Furthermore the KNN model showed high rank order correlation with the RandomForest, MLP and XGBoost model. Notably, not displayed in Fig. 2 the PSD slope features showed the highest group importance for all models, highlighting the relevance of PSD features to brain-age prediction.

## VIII. REVISITING PREVIOUS FINDINGS

In this section we attempt to validate hypotheses from prior research (Outlined in Table I) with our findings derived from our *SHAPagreement* analysis.

Notably, previous research emphasizes a decrease in delta and theta activity during the aging process [1], [5]. While mean absolute power did correlate with higher age prediction in most models, more informative features were the kurtosis and quantiles, indicating a "tighter" distribution of theta and delta power features for older subjects, see Fig. 5 a) and b). Moreover, Vandenbosh et al. found that central channels were indicative of age [1]. However, in all models features from frontal and temporal signals had the highest SHAP values (see Fig. 4 and appendix section II).

The decline in low-frequency bands is a recognized indicator of brain maturation in children and adolescents [7], [8]. This notion is strongly supported by our analysis, especially in the ElasticNet and random forest models. Once more, the converging trends of the upper 75% and lower 25% quantiles pointed towards higher age predictions. Furthermore, a noticeable decline in the intercept of the log-log PSD regression line within the theta band correlated with higher age predictions, as illustrated in Fig. 5 j).

Among alpha Band features, signal power features had the highest SHAP values, replicating previous results [1], [5]. For instance, increase in power was highly indicitive in higher age prediction in XGBoost and CatBoost (see Fig. 5 d). Additionally, zero crossing and higuchi fractal dimension values (Fig. 5 h) within alpha band exhibited substantial correlation with higher age predictions, suggesting greater alpha band signal complexity in older subjects. Regional alpha band importance also aligned with previous results, with important features mostly clustering in frontal and parietal electrodes [1]. While [1] used random forest, we found similar alpha importance patterns in the SVR, KernelRidge and MLP models (Fig. 4).

Beta band features had the lowest aggregated SHAP scores across all models, indicating that changes between early ages (5 to 21 in our data) mostly happened in low frequency bands. Nevertheless, beta band power was informative to both the XGBoost and CatBoost models (Fig. 5 k)). Additionally, in the SVR model the beta band kurtosis also correlated with higher age predictions, implying that over all beta activity increases with brain maturation. These findings align with previous work [1], [2], [5]

Van Noordt and Willoughby found that power spectral density measures are informative for brain-age detection [7]. Indeed, all models trained on set 12-All found spectral slope features as task relevant. Specifically, we found that younger subjects had higher R2 coefficients. Finally, like Al Zoubi et al [2], we also found that spectral entropy is positively correlated with the brain age (see Fig. 5 c).

Generally, while most results identified in our literature review were replicated, a few were not. Moreover, while we found high variability between different groups of models (for instance regression vs decision trees, see Fig. 2), most replicated hypotheses could be reproduced by the vast majority of our models. Indicating that models agree on core concepts and that disagreements, while frequent, are inconsistent. We conclude that productive research should focus on features consistently identified as predictive by various algorithms, rather than patterns uncovered by only the best performing model.

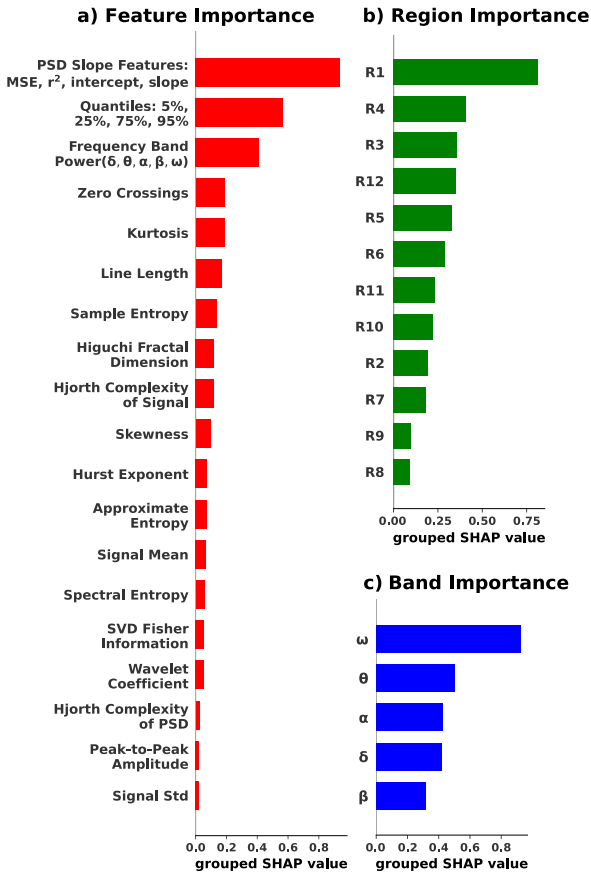

Fig. 3: Ranked group importance as sum over absolute SHAP values (Model: XGBoost, Training Set: 12-All). Features are grouped by a) method/measure b) electrode location c) frequency band.

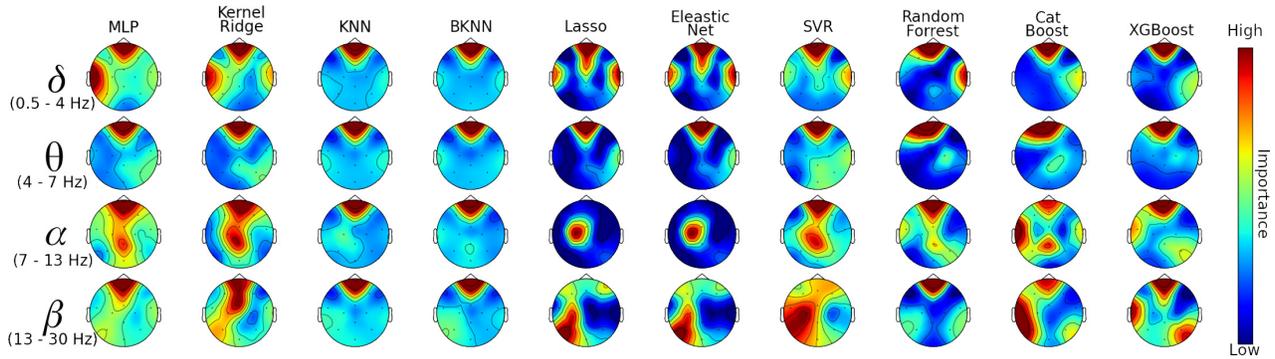

Fig. 4: Each model was trained on set 12-All. Columns show the regional importance for one model while the row varies the frequency band. The color coding indicates the normalized regional SHAP importance picked up by the model.

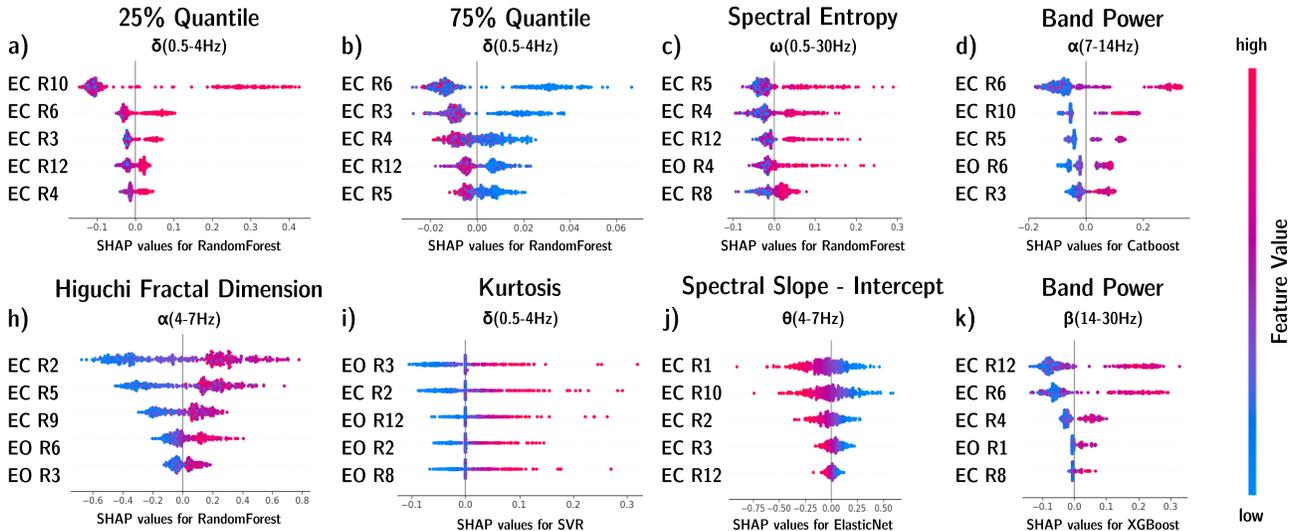

Fig. 5: SHAP values (impact of a feature on the model's prediction) computed for RandomForest, CatBoost, XGBoost, SVR and ElasticNet on 12-All. Each plot is specific to a feature in a frequency band: a) $\delta$ 25% Quantile b) $\delta$ 75% Quantile c) $\omega$ Spectral Entropy d) $\alpha$ Band Power e) $\alpha$ Higuchi Fractal Dimension f) $\delta$ Kurtosis g) Intercept of the log-log PSD regression line in $\theta$, h) $\beta$ Band Power. Labels indicate the resting state (EC: eyes closed, EO: eyes open) and region (see Fig. 1). Color coding represents the feature value while x-axis depicts shap values. Each point corresponds to a testing sample.

## IX. FEATURE SELECTION IN DEEP LEARNING

Machine learning research has been steadily moving towards black-box data driven deep learning approaches. These methods are especially problematic in bio-informatics and medical applications that emphasize interpretability [37], [38]. Recent studies suggest the alternative of combining expert feature-handcrafting and deep-learning. Here we follow the Feature Imitating Networks (FINs) framework proposed by [38], which achieved state of the art results in speech recognition, MEG and EEG decoding, and financial forecasting [38], [39]. FIN are ensembles of deep learning models, trained on raw EEG data to predict specific underlying statistical features known to be relevant in an important field. Leveraging the principals of transfer learning, a meta-learner is attached to the pre-trained models, harnessing the learned feature space to tackle novel tasks. In our study, we trained FIN models to forecast the upper (75%) and lower (25%) quantiles, using EEG signals extracted from five distinct frequency bands ($\delta, \theta, \alpha, \delta, \omega$). The FIN achieved performance comparable to the best other models (See appendix for full training details and breakdown of the results). Overall, this demonstrates that, despite constant advancements in machine learning, feature selection is likely to remain an integral aspect of bio-informatics research. See supplementary section III for FIN implementation and analysis details.

## X. CONCLUSION

Replication is a fundamental aspect of scientific research. This paper highlights the risk of not examining the robustness of scientific hypotheses across different machine learning algorithms. Our study, focusing on brain-age prediction using EEG data, demonstrates significant variations in outcomes based on the choice of algorithm. This finding underscores the necessity of diversifying algorithm selection in research to avoid narrow interpretations.

Contrasting our work with existing studies, we emphasize the importance of using a range of algorithms, beyond just the best-performing ones. The challenges identified, particularly in interpretability in deep learning, call for future research to establish more rigorous practices in machine learning application. We hope our findings encourage the research community to adopt comprehensive testing and replication strategies, ensuring more reliable and robust scientific outcomes. Such an approach is vital for advancing our understanding of complex biological data through machine learning.

# Supplementary Material: Different Algorithms *(Might)* Uncover Different Patterns

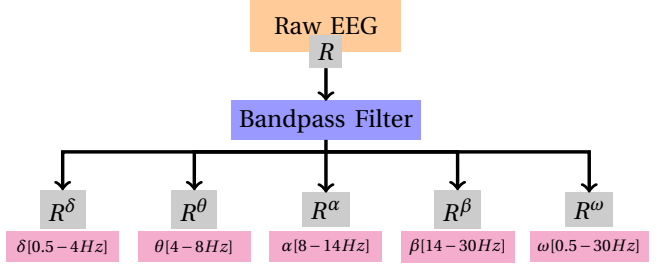

Fig. A.1: Illustration on how a EEG recording is filtered into the frequency bands of interest.

## I. Feature Extraction Pipeline

Our feature extraction pipeline's architecture closely aligns with the procedural framework outlined in the work of [1], [2]. The focal frequency bands for our feature extraction encompass delta (0.5 - 4 Hz), theta (4 - 7 Hz), alpha (7 - 14 Hz), beta (14 - 30 Hz), as well as the whole spectrum $\omega$ (0.5 - 30 Hz) encompassing all frequency bands. To initiate the process, we implement a band-pass filter on recording $R$ for each distinct frequency band. This yields filtered recordings denoted as $R^{fb}$, where $fb \in \delta, \theta, \alpha, \beta, \omega$ defines the frequency band specific filter applied on $R$ (see Fig. A.1). Following this step, each signal $R^{fb}$ is divided into a sequence of epochs (see Fig. A.2).

Lets denote a trail of $n \in \mathbb{N}$ epochs obtained by splitting a recording $R^{fb}$ as $E^{fb} = (R_1^{fb}, R_2^{fb}, ..., R_n^{fb})$. In this representation, each epoch $R_i^{fb} = (c_{i1}, ..., c_{im})$ consists of a collection of channels denoted as $c_{ij} = (t_k, t_{k+1}, ..., tk + \Delta s)$. Here, $\Delta s$ is defined as the product of the *sampling rate* and the *epoch duration*, encapsulating a span of time points.

We set the epoch duration to 2s for eyes open and to 4s for eyes closed to obtain 10 epochs from each recording. For feature extraction we used the *extract_features* method from the *mne_features* python library [3].

Let us symbolize a sequence of features derived from a series of epochs $E^{fb}$, as $F^{fb} = (F_1^{fb}, F_2^{fb}, ..., F_n^{fb})$. Within this context, the notation $F_i^{fb} = (f_{i1}, f_{i2}, ..., f_{im})$ takes on the role of representing the unique features intrinsic to an epoch $R_i^{fb}$, whereas $f_{ij}$ represents a vector that captures a range of different features extracted from the signal $c_{ij} \in R_i^{fb}$ from the $i$-th epoch in channel $j$.

After extracting the sequence of features $F^{fb}$ from a trail of epochs $E^{fb}$, we averaged the feature vectors across all epochs, given by $\overline{F}^{fb} = (\overline{f_1}, \overline{f_2}, ..., \overline{f_m})$ with $\overline{f_j} \in \overline{F}^{fb}$ being the mean feature vector over all $n$ epochs from channel $j$.

$$\overline{f_j} = \frac{1}{n} \sum_{i=1}^{n} f_{ij}$$

The averaged trail of features $\overline{F}^{fb}$ for each frequency band $\delta, \theta, \alpha, \beta$ and $\omega$ is than flattened to a one dimensional vector and represents one training sample.

## II. Further SHAP Analysis

For our top-performing model, XGBoost, the SHAP values indicated that higher slope values of the log-log PSD regression line, specifically in the omega band (0.5-30Hz), were positively correlated (refer to Fig. B.3 f)). This suggests a shift: as aging occurs, there's a decrease in lower frequencies and an increase in higher frequencies. The R2 coefficient of the log-log PSD regression fit exhibited significant model impact, as depicted in Fig. B.3 c). However, translating these results into actionable insights demands deep domain expertise. As highlighted in the primary section of the paper, aging led to a convergence between the upper 75% and lower 25% quantiles of theta activity. This trend is evident in Fig. B.3 a) and b). In the beta band, a reduced kurtosis with higher age predictions hints at a wider signal distribution around the mean (see Fig. B.3 e). Meanwhile, elevated Zero Crossing values in the alpha band, as shown in Fig. B.3 d), suggest heightened neuronal activity in this frequency range with aging.

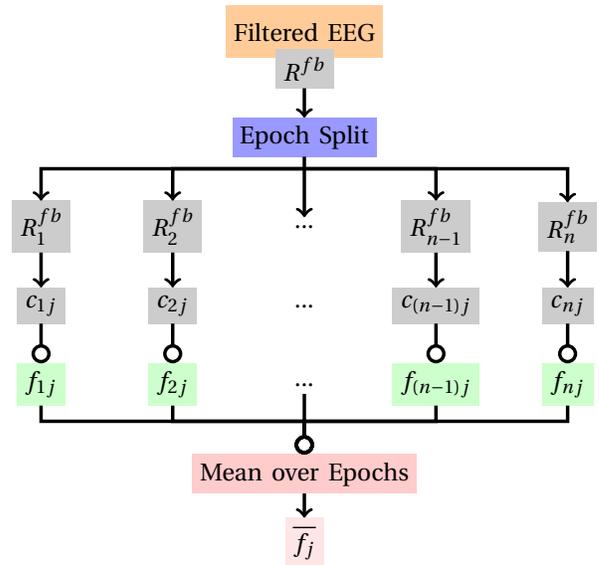

Fig. A.2: Illustration on how a filtered EEG signal is split into a set of $n$ epochs. The feature vector $f_{ij}$ is extracted from the epoch channel $c_{ij}$. By averaging over the feature vectors $f_{ij}$ of an epoch $i$, we obtain $\overline{fj}$.

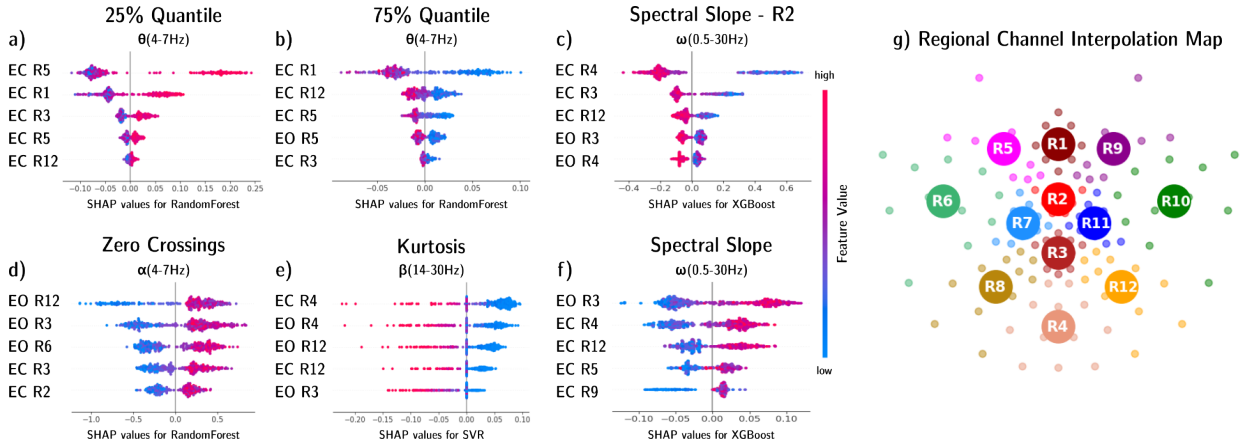

Fig. B.3: SHAP values (impact of a feature on the model's prediction) computed for RandomForest, XGBoost and SVR on 12-All. Each plot is specific to a feature in a frequency band: a) 25% and b) 75% Quantile in theta, c) R2 coefficient of the log-log PSD regression line fit, d) Zero Crossings in alpha, e) Kurtosis in beta, f) Slope of the log-log PSD regression line in omega. Figure g) shows a topological map of the sensory locations (subject facing up), where Small dots show electrodes and big dots the regional electrode. Labels on the y-axis indicate the resting state (EC: eyes closed, EO: eyes open) and region. The color coding represents the feature value, while the x-axis depicts its impact on the model's prediction. Each point corresponds to a testing sample.

## III. FEATURE IMITATION NETWORK

To train our Feature Imitation Network (FIN) models, we incorporated 400 non-labeled subjects from our dataset, originally designated for the brain age prediction challenge [4]. As these subjects couldn't be used for training age prediction models, they were, however, invaluable for training our FINs, which aim to predict underlying signal features. This approach expanded our training set to 3085 recordings, post-elimination of 115 faulty entries, a process detailed in the main body of the paper. We subjected all EEG recordings to the identical preprocessing routine previously established in our feature extraction pipeline, ensuring consistency across our dataset. Additionally, to simplify the FIN ensemble's overall complexity, we took a strategic step to further reduce channel dimensionality. We grouped the electrodes into five distinct regions, as depicted in Figure C.4, and performed signal averaging to interpolate the channels. Each recording was segmented into 2-second epochs (500 time points) after band-pass filtering to highlight our targeted five frequency bands $(\delta, \theta, \alpha, \delta, \omega)$. These filtered epochs formed individual samples in our training set. Using the *mne_features* library, we extracted the 75% and 25% quantiles from these signals [3], which became the target metrics for our FIN models.

We trained FIN models on EEG signals from five frequency bands $(\delta, \theta, \alpha, \delta, \omega)$ to predict the 75% and 25% quantiles. Inputs for each FIN model were one-dimensional vectors, each representing five regionally interpolated chan-nels from an epoch, with a total of 2,500 features. The outputs were 10-dimensional vectors, based on the two quantiles forecasted for each channel. These models were trained on data specifically filtered for each frequency band, resulting in five unique FIN models. Refer to Figure C.5 for a visual representation of a FIN model. After every layer, except the output, we incorporated batch normalization, applied an activation function, and introduced dropout. For optimal hyper-parameters, we conducted a randomized grid search, experimenting with aspects like the number and size of hidden layers, learning rate, batch size, activation function, dropout rate, weight decay, and optimizer momentum. Additionally, we employed the *MinMaxScaler* from the scikit-learn package [5] to scale the target data during training. In Table I you can see the over all performance results for each of our five fin models. You can find the scatter plots of the actual and predicted values, as well as the error distribution plot in Figure C.7 and Figure C.8.

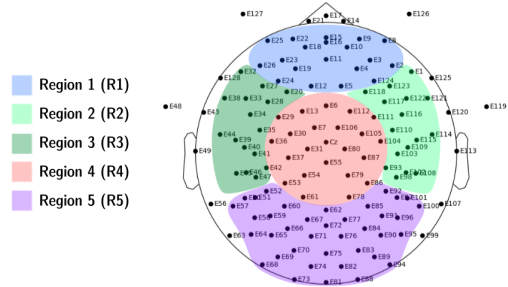

Fig. C.4: Topological map of the electrodes grouped into 5 regions. Each color represents a group of electrodes which were combined through signal averaging in effort of reducing the channel dimension.

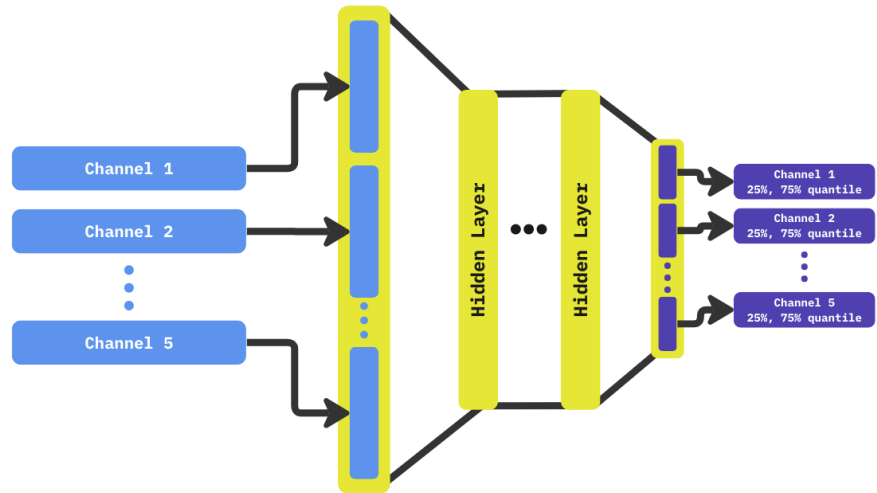

Fig. C.5: The figure illustrated one FIN model for predicting the upper 75% and lower 25% quantile for all five input channels specifically filtered into a distinct frequency band.

TABLE I: The table shows the $R^2$ regression score each model trained for a specific frequency band achieved.

| FIN Model | $R^2$ |
|---|---|
| quantile (delta) | 0.863 |
| quantile (theta) | 0.761 |
| quantile (alpha) | 0.930 |
| quantile (beta) | 0.946 |
| quantile (omega) | 0.842 |

We assembled the FIN ensemble by removing each model's output layer and directly connecting its final layer to the meta-learner's input, as illustrated in Figure C.6. Initially, the FIN models were frozen, with only the meta-learner's weights adjustable during training. Once this phase concluded, we unfroze the FIN models for end-to-end training to refine the entire ensemble. Hyper-parameter tuning was executed for the meta-learner, which shared the same architectural blueprint as our FIN models. This tuning process mirrored the adjustments we made during the FIN training phase.

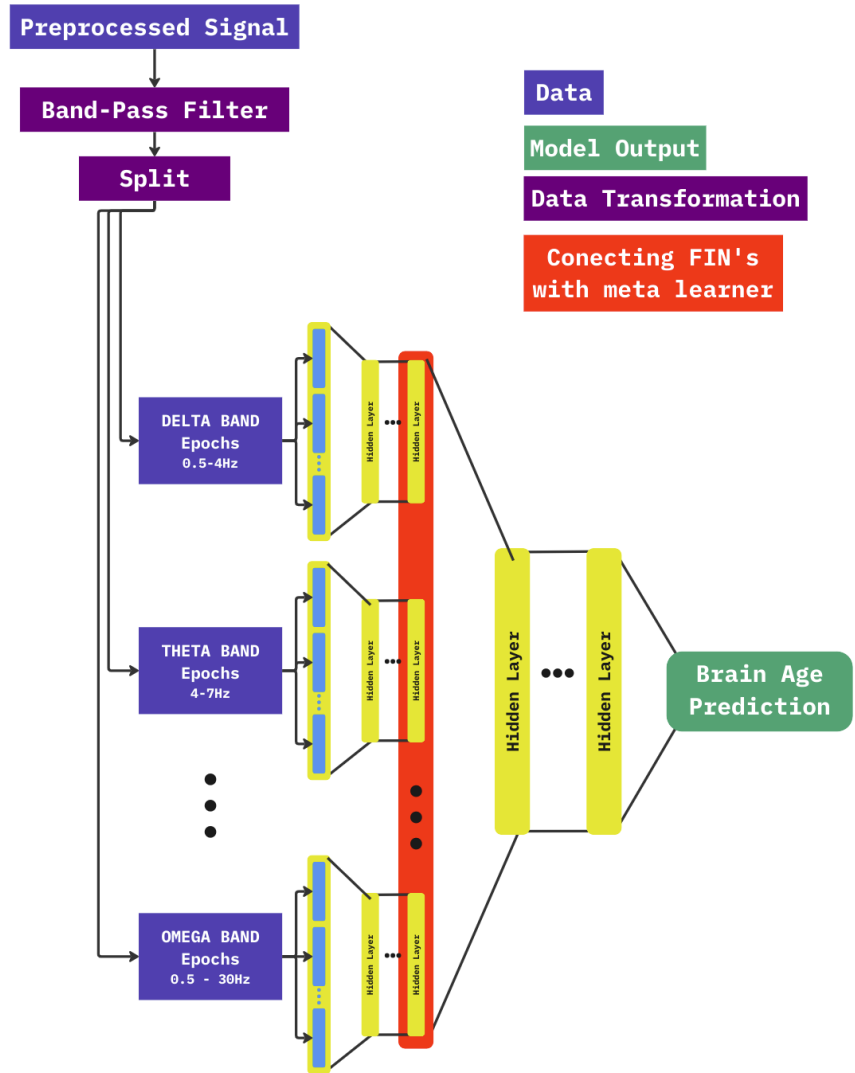

Fig. C.6: The figure illustrates the ensemble of FIN models all connected to the meta-learner predicting the age. For each FIN model, the output layer is removed. The last layer of the FIN models is then connected to the input layer of the meta-learner.

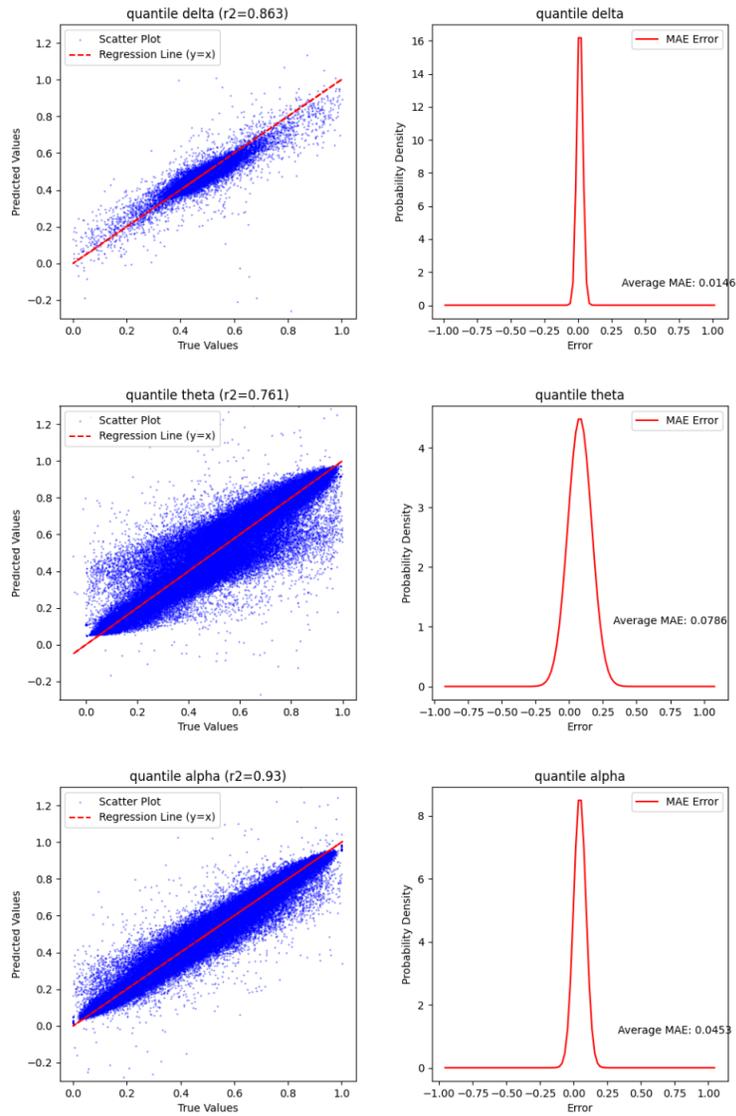

Fig. C.7: The figure shows scatter plots for the actual quantile values and the predicted values for the FIN trained on delta (top), theta (middle) and alpha (bottom) filtered data.

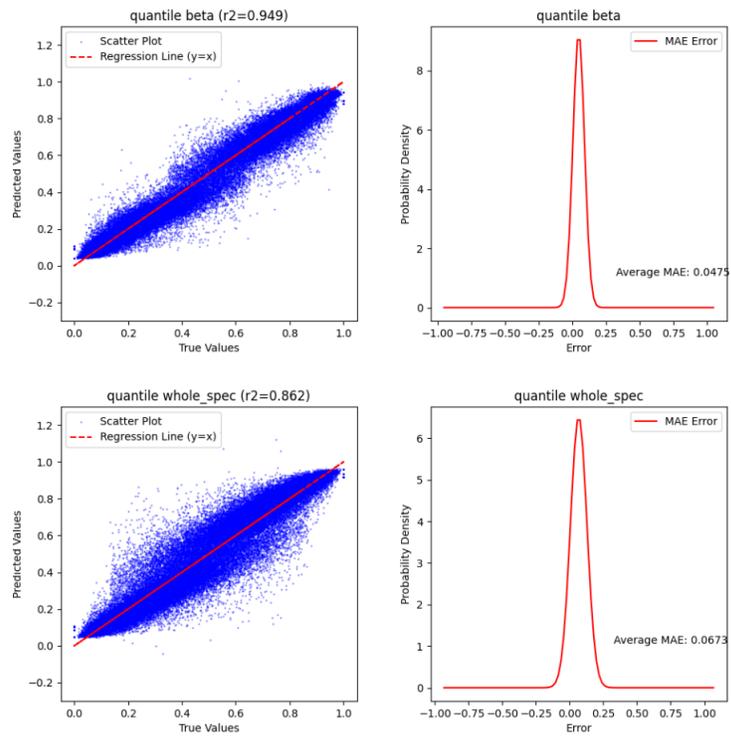

Fig. C.8: The figure shows scatter plots for the actual quantile values and the predicted values for the FIN trained on beta (top) and omega (bottom) filtered data.